\documentclass{elsart}

 \usepackage{epsfig}

\usepackage{amssymb}

\begin{document}

\def\NZ{{\bar N_0}}
\def\NJ{N_{\J}}
\def\J{J/\psi}
\def\JM{$J/\psi~$}
\newcommand{\Npart}{N_{part}}
\newcommand{\Nc}{N_c}
\newcommand{\Nch}{N_{ch}}
\newcommand{\Ncbar}{N_{\bar c}}
\newcommand{\ccbar}{c \bar c}
\newcommand{\cbar}{\bar c}
\newcommand{\Nccbar}{N_{\ccbar}}
\newcommand{\kts}{\langle{k_T}^2\rangle}
\newcommand{\pts}{\langle{p_T}^2\rangle}
\newcommand{\epsz}{\epsilon_0}
\newcommand{\sigF}{\sigma_F}
\newcommand{\sigD}{\sigma_D}
\def\pt{$p_{T}~$}
\def\RAA{$R_{AA}~$}
\newcommand{\gev}{\mathrm{GeV}}

\begin{frontmatter}


 \title{Quarkonium production via recombination}
 \author{R. L. Thews}
 \address{Department of Physics, University of Arizona, Tucson, AZ 85721 USA}


\begin{abstract}
The contrast between model predictions for the transverse momentum spectra
of \JM observed in Au-Au collisions at RHIC is extended to
include effects of nuclear absorption.  We find that the difference
between initial production and recombination is enhanced in the
most central collisions.  Models utilizing a combination of these
sources may eventually be able to place 
constraints on  their relative magnitudes. \end{abstract}


\end{frontmatter}

\section{Introduction}
\label{intro}

The original motivation \cite{Thews:2000rj} to consider heavy quarkonium production in a
region of color deconfinement was the realization that the
number of charm pairs initially produced in a central heavy collision
will be at least of order 10 
\cite{Adams:2004fc,Adler:2004ta} at RHIC 200 energy. Preliminary
model estimates indicated that interactions of a quark and antiquark which
were not produced in the same nucleon-nucleon collision (the so-called
off-diagonal pairs) will result in formation of additional 
\JM (now called recombination)
which would mask the expected suppression \cite{matsuisatz} due to 
medium effects.
However, these model estimates proved to be very sensitive to details
of various parameters which control 
the recombination process \cite{sqm2003}.
PHENIX
results have now provided a first look at the total \JM population
in terms of the ratio \RAA over a large range of collision centrality \cite
{PereiraDaCosta:2005xz}.  
Surprisingly, one sees a suppression factor which is very close to that
observed at SPS energy.  A number of models predicted that there
would be substantially more suppression at RHIC, and all were thus
incompatible with these initial results.  However, models in which
some sort of in-medium recombination process is included find a total
\JM population which is not incompatible with the PHENIX results.  
This can be taken
as evidence that the recombination process is indeed significant at
RHIC energy.  However, one must have confidence that the suppression
parameters used are indeed appropriate at RHIC, which at this time
has no independent confirmation.  In fact, there is a proposed alternate 
scenario \cite{Satz:2005hx,Karsch:2005nk}
in which
the observed suppression at both SPS and RHIC energies is entirely due to
dissociation of the $\psi^{\prime}$ and
$\chi_c$ which would otherwise have provided additional \JM from decay 
feed-down.
\vspace{0.5truecm}
\begin{figure}[htb]
\epsfig{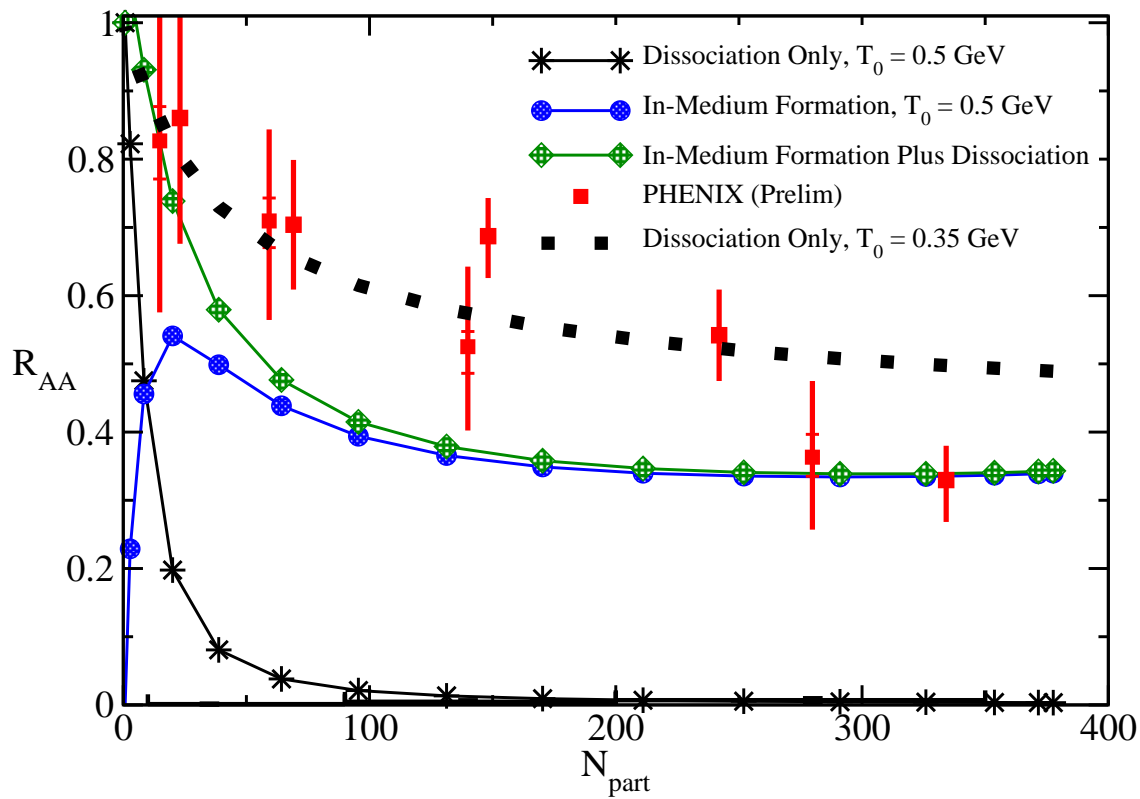}
\caption{\small Comparison of dissociation and recombination model predictions
for \JM \RAA.}
\label{RAAplot}
\end{figure}
\vspace{0.5truecm}
We show in Figure \ref{RAAplot} some model predictions with
and without recombination compared with the initial PHENIX measurements.
The set of model parameters was motivated in \cite{sqm2003} which uses 
an initial temperature $T_{0}$ = 500 MeV and $\Nccbar = 10$ initially-produced
$\ccbar$ pairs for central collisions, where the charm quark momentum
distribution is 
calculated in pQCD.  The prediction for
no recombination (stars) leads to almost complete suppression of 
initially-produced \JM.  Inclusion of in-medium charm quark recombination
(blue circles) dominates over almost the complete centrality range, and the sum
(green diamonds) is seen to follow the trend of the measured points 
fairly well.  We
note that since the normalization of \RAA was not contained in the model
predictions, we choose to normalize \RAA to unity at the most
peripheral points.

To illustrate the fragility of such an interpretation, we also show
model predictions  in which no recombination process occurs and the initial
temperature is decreased to 350 MeV.  The resulting
 \RAA (black squares)
is also seen to be roughly compatible with the data.  Thus in the absence of
independent information on the initial temperature parameter in this
model, we cannot use the \RAA measurements to make a definitive 
statement.

Fortunatately, the transverse momentum spectra of the
surviving \JM are also measured, and we can use this additional
information to test the various scenarios. 
In the next section we review the use of spectra as a signal of
the production mechanism.  The following section summarizes the crucial 
role of centrality dependence and presents a comparison with the
PHENIX measurements.  Finally, we include for the first time 
the effects of nuclear
absorption on these signals, and illustrate the experimental
precision necessary for a definitive result. 
\section{Spectra as signals of recombination}
\label{spectra}

The number of \JM formed in a medium via recombination of pairs of
$c$ and $\cbar$ quarks is determined by the competition between the
recombination and dissociation cross sections.  In the case where
the total number $\NJ$ is much less than the 
initial number of quark pairs $\Nccbar$, the solution is of the form

\begin{equation}
\NJ(t_f) = \epsilon(t_f) [\NJ(t_0) +
\Nccbar^2 \int_{t_0}^{t_f}
{\lambda_{\mathrm{F}}\, [V(t)\, \epsilon(t)]^{-1}\, dt}],
\label{eqbeta}
\end{equation}
where $t_0$ and $t_f$ define the lifetime of the deconfined region,
and $\lambda_{\mathrm{F,D}}$ are the formation and dissociation 
cross section reactivities, respectively, and $V(t)$ is the
(expending) volume of the deconfinement region.  
Note that the function $\epsilon(t_f) = 
e^{-\int_{t_0}^{t_f}{\lambda_{\mathrm{D}}\, \rho_g\,
dt}}$
would be the suppression factor in this scenario if the
formation mechanism were neglected.

One sees that the second term, quadratic in $\Nccbar$, is precisely the
total number of possible recombinations which 
could occur in the deconfinement volume,
modified by the factor $\epsilon(t_f)/\epsilon(t)$, which is just the
suppression factor for \JM  formed between times $t$ and $t_f$. 
To calculate the transverse momentum or rapidity spectra of the
\JM formed via recombination, one must use a version of this solution
which is differential in the \JM momentum.  Since the cross section
used here is simple gluo-ionization of the \JM bound state by the
interaction of its color dipole with free gluons in the medium, the
differential cross section expression is trivial.  To solve the
equation, we first generate a set of $\ccbar$ pairs according to pQCD
amplitudes \cite{mangano} to fix the kinematic distributions 
of the recombining quarks.
The resulting spectra of the \JM is then

\begin{equation} 
\frac{dN_{\J}}{d^3 P_{\J}} = \int{\frac{dt}{V(t)}}
\sum_{i=1}^{N_c} \sum_{j=1}^{N_{\bar c}} {\it {v}_{rel}} 
\frac{d \sigma}{d^3 P_{\J}}(P_c + P_{\bar{c}} \rightarrow P_{\J} + X),
\label{formdist}
\end{equation}  where the sum over all $\ccbar$ pairs incorporates
the total formation reactivity. 

The charm quark spectra we calculate from pQCD have used
collinear parton interactions only.  To simulate the effects of
confinement in nucleons, and more importantly the initial state
\pt broadening due to multiple scattering of nucleons, we have
supplemented the quark pair \pt distributions by adding a
transverse momentum kick to each quark in a diagonal pair,
chosen from a Gaussian distribution with width characterized by
$\kts$.  This effect is magnified in the difference between diagonal and
off-diagonal pairs.  Since the azimuthal direction of the $\vec{k}_T$
is uncorrelated from pair-to-pair, an off-diagonal pair with initial
$\pts$ will inherit an increase of  2 $\kts$.  However, the azimuthal correlation
inherent in the diagonal pair will result in an increase 
4 $\kts$. This effect is evident when we investigate the $\kts$ dependence
of charm quark pairs and \JM formed from recombination. We find that:
\begin{itemize}

\item[] ${\langle{p_T}^2\rangle}_{diagonal~ccbar}$~= 0.8 + 4 $\kts$ GeV$^2$  
\item[] ${\langle{p_T}^2\rangle}_{off-diagonal~ccbar}$~= 4.9 + 2 $\kts$ GeV$^2$  
\item[] ${\langle{p_T}^2\rangle}_{in-medium~\J~formation}$~= 2.4 + $\kts$ GeV$^2$  
\end{itemize}

We use reference data on 
$\J$ production in pp interactions \cite{Adler:2005ph}, which
of course involves diagonal pairs, to fix $\kts_{pp} = 0.5 \pm 0.1$ GeV$^2$.
The corresponding value in nuclear interactions depends on the centrality,
which we determine in the next section.

\section{Centrality Dependence of $\pts$ }
\label{}
We use the standard random walk picture of initial state NN interactions
to model the transverse momentum broadening in pA and AA interactions
\cite{Gavin:1988tw}.
This uses a parameter $\lambda^2$, which is {\it proportional} 
to the average 
$\kts$ per initial state collision (a factor
of 0.5 times the probability for initial state parton elastic
scattering in a given inelastic NN collision is conventionally absorbed 
in the definition).
 The net broadening in a pA collision
can be parameterized in terms of 
the product of this $\lambda^2$ and $N_n$, the average number of inelastic
collisions each nucleon incurs {\it before} the interaction in which the
charm quark pair is produced.  In terms of the total number of collisions
per initial state nucleon (n), 
this number is 
\begin{equation}
N_n = \frac{1}{n}\sum_{m=1}^n (m-1) = \frac{(n-1)}{2}.
\label{collisionpoint}
\end{equation}

\begin{figure}[htb]
\epsfig{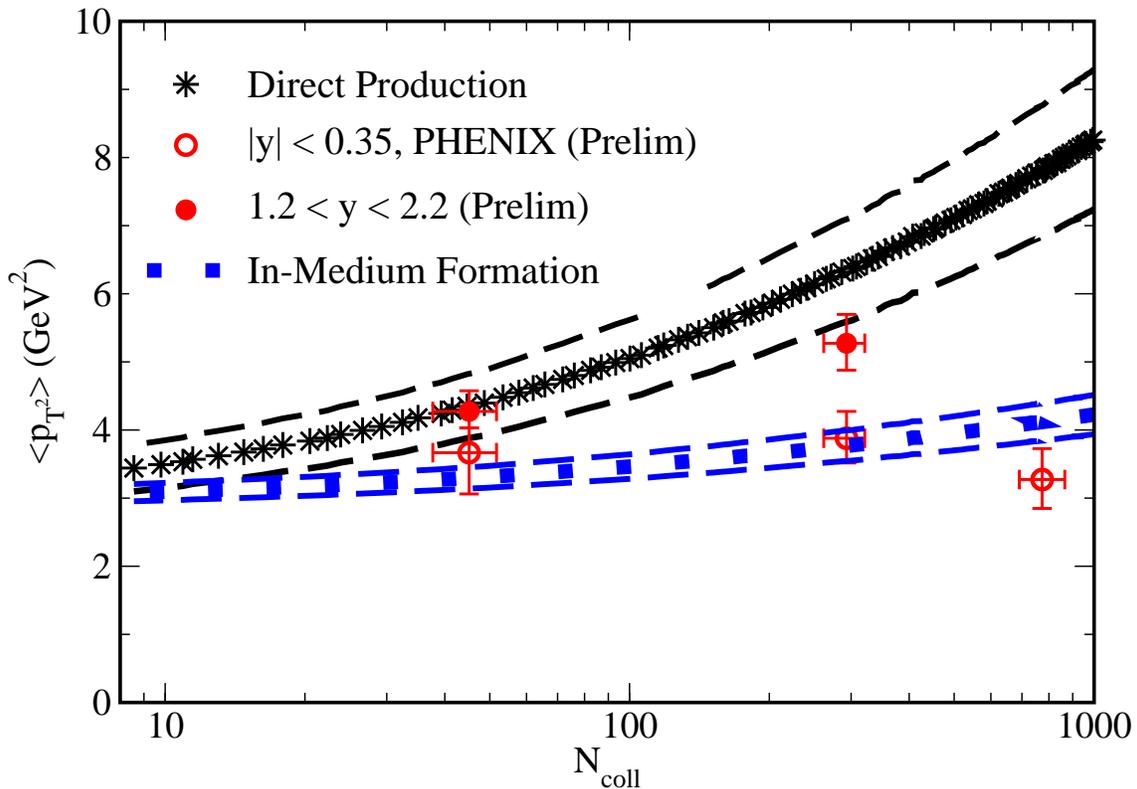}
\caption{\small Centrality dependence of $\pts$ comparing
model predictions for initial (direct) production and in-medium formation, 
both assuming  charm quarks with momentum spectra as predicted by pQCD. 
}
\label{directvsformationpt2}
\end{figure}

We extract $\lambda^2 = 0.56 \pm 0.08$ GeV$^2$ 
from the nuclear broadening
measurements of \JM produced in minimum bias d-Au interactions
\cite{Adler:2005ph}.  The corresponding $\kts_{AA}$ is then
calculated as a function of centrality, parameterized by the
total number $N_{coll}$ of binary collisions.  The final results
for $\pts$ of initially-produced \JM from diagonal $\ccbar$ pairs
and $\pts$ of \JM formed in-medium are shown in Figure \ref{directvsformationpt2}.  
One sees that they differ by increasing amounts as centrality
increases, which is a direct consequence of their different
$\ccbar$ pair precursors (diagonal or off-diagonal).  We note that
the initial production curve (higher $\pts$) should describe the
sequential production scenario, in which all observed \JM originate
from unsupressed direct production \cite{Karsch:2005nk}.  The in-medium
production curve (lower $\pts$) would describe a situation in which
all of the initially-produced \JM have been screened away, and thus
all observed \JM are due to in-medium formation processes. 
We should stress that these calculations of $\pts$ use the charm
quark momentum distributions which follow from initial production
via pQCD amplitudes.
 One might
expect that the actual physical situation would be some weighted
sum of these two extreme cases. (For a 
discussion of comparison with preliminary data, see \cite{sqm2006}.)
We should stress that these calculations of $\pts$ use the charm
quark momentum distributions which follow from initial production
via pQCD amplitudes.  
\vspace{0.5truecm}
\vspace{0.5truecm}

\section{Effects of Nuclear Absorption}

There is an additional effect on the \pt spectrum of 
directly-produced \JM due to 
their interaction in cold nuclear matter which they encounter
on their path out of the interaction region. We note that
this effect cannot modify the \JM formed via recombination
in the medium, since the formation process takes place at
a later stage. 
One could include the nuclear absorption effect in  
a full Glauber calculation of the initial production
of \JM, but for simplicity we here extract a correction factor for
the \pt spectra using an effective absorption cross section in
a path length geometry.  This will involve an extension of the
random-walk picture of individual NN collisions which occur both
before and after the point at which a hard interaction produces the
$\ccbar$ pair which eventually emerges as a $\J$.  What we wish to 
calculate in this picture is the average number of such NN collisions
which occur {\it before} the collision in which the $\J$ production is
initiated.  In the absence of nuclear absorption, the relative
probability of observing a $\J$ will be independent of the
spatial position of its production point. This leads to the 
expression in Eq. \ref{collisionpoint}.  
With nuclear absorption, however, one must weight each of the
possible production points by the probability that the $\J$
produced at a given position will survive to be observed in
the final state.

Let $P_{mn}$ be the relative probability that the $\J$ survives its
path through the remaining nuclear matter, where n is the total
number of NN interactions and m specifies the number in this
sequence at which the $\J$ production point occurs. We parameterize 
this in a simple straight-line trajectory model, to obtain

$P_{mn} = {{\epsilon}_{n}}^{n-m}$, where $\epsilon_n = exp(-\rho 
\sigma L_{max}/n)$ is
the average absorption factor encountered by the $\J$ in each path
length between NN collisions.  Here $\rho$ is the density of nuclear
matter and $\sigma$ is the absorption cross section.

Then the average effective number of NN collisions which an
{\it observed} $\J$ will have experienced is 

\begin{equation} 
N_n = \frac{1}{P_{tot}} {\sum_{m=1}^{n} (m-1) P_{mn}},
\end{equation}
where
\begin{equation}
P_{tot} \equiv \sum_{m=1}^{n} P_{mn} = \frac{{\epsilon_n}^n - 1}{ 
{\epsilon_n}-1}.
\end{equation}

\begin{figure}[htb]
\epsfig{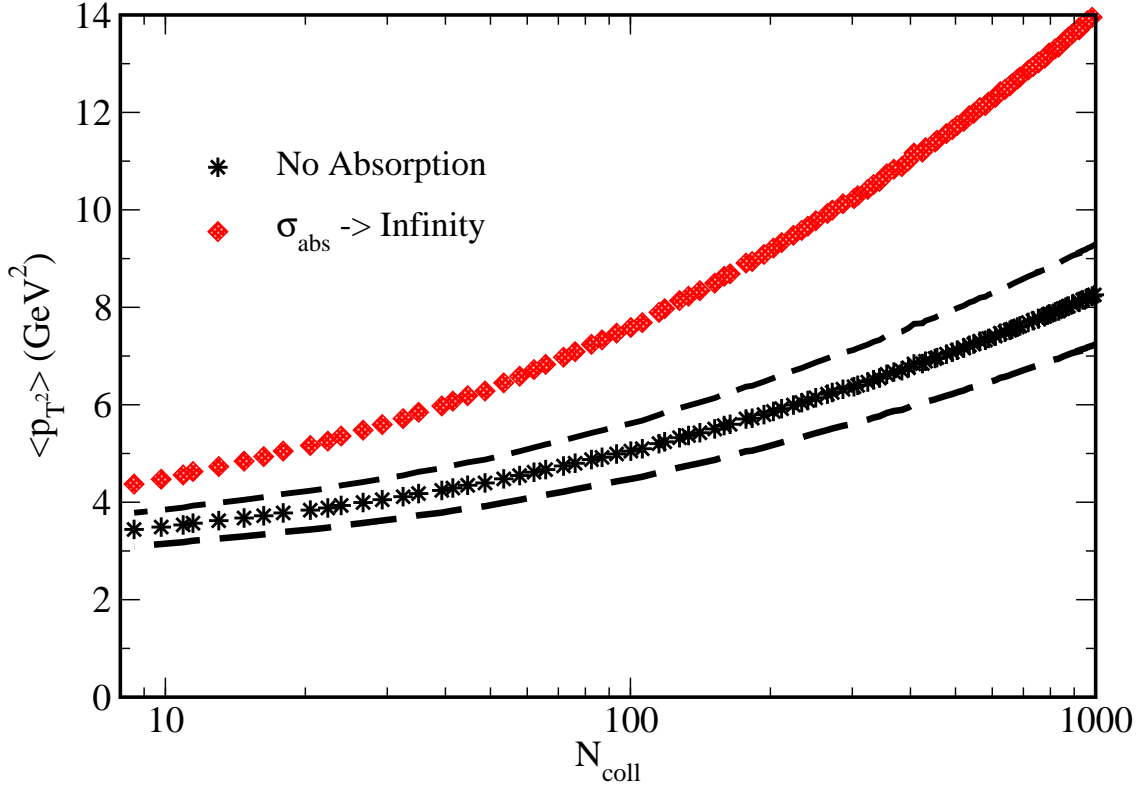}
\caption{\small Maximum increase of $\pts$ for initial (direct)
production in the limit of infinite nuclear absorption.} 
\label{infiniteabs}
\end{figure}

In the limiting case of zero nuclear absorption, $P_{mn}$ = 1, $P_{tot}$ = n, 
and one recovers $N_n = \frac{n-1}{2}$.  In the opposite limit of 
infinite absorption, $P_{mn}$ vanishes unless m=n, resulting in
$N_n$ = n-1.  This correctly describes the effective number of
initial state 
collisions for a $\J$ which is produced in the final collision point
at the back of the nucleus, the only possibility which avoids
complete absorption.  We show in Figure \ref{infiniteabs}
the corresponding (maximum) increase in $\pts$.

\begin{figure}[htb]
\epsfig{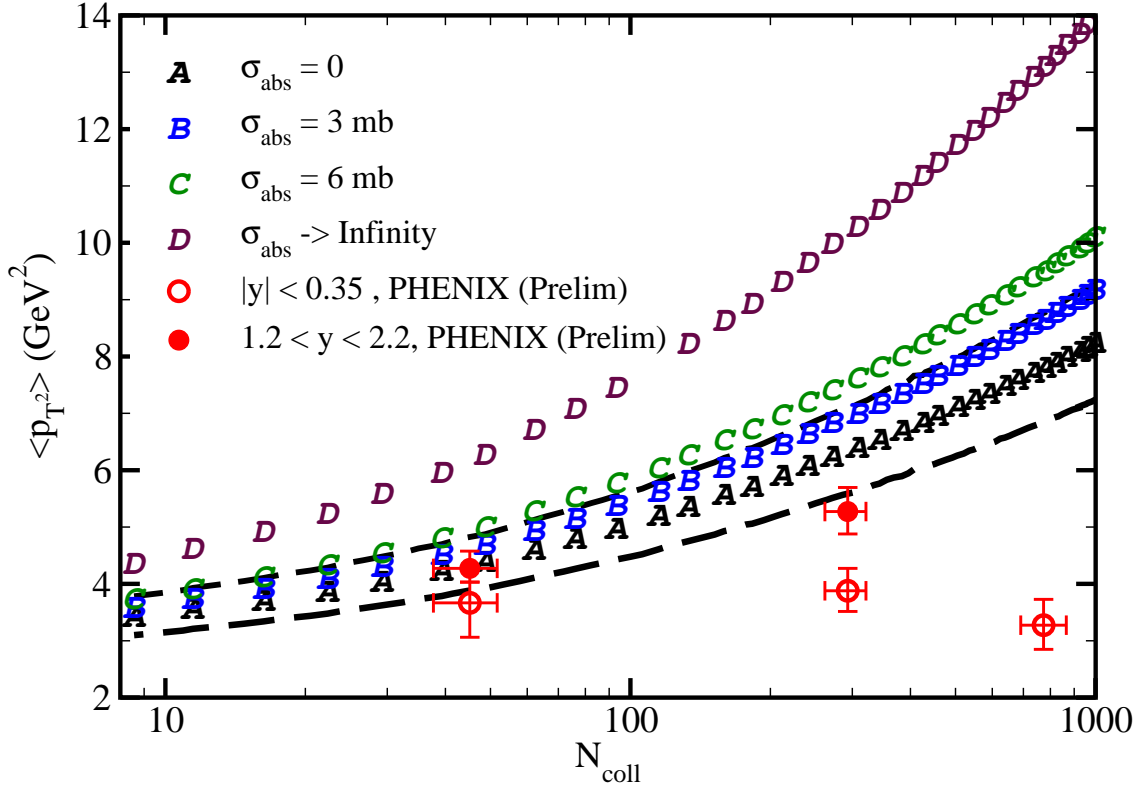}
\caption{\small Variation with nuclear absorption of $\pts$ 
for initial (direct) production.}
\label{pt2andnucabs}
\end{figure}

\vspace{-0.5truecm}
For finite nuclear absorption in terms of
 $\beta \equiv \rho~\sigma~L_{max}$,
a numerical evaluation  
yields $N_n \approx (1 + 0.2~\beta)~\frac{n-1}{2}$, for
$0~\leq~\beta~\leq 1.5~$.
We show in Figure \ref{pt2andnucabs} the predicted $\pts$
 spectra of
initially produced diagonal $\ccbar$ pairs, as appropriate for
initial production of $\J$ in Au-Au interactions.  One sees that
the effect of nuclear absorption always results in increased 
$\pts$, which
enhances the difference with respect to in-medium $\J$ production.
Even the smallest absorption cross section considered (3 mb) shifts the
prediction upward to an extent which may prove difficult to
reconcile with preliminary PHENIX data at both forward and mid rapidity
\cite{sqm2006}.
\vspace{0.7truecm}



\end{document}